\documentclass[useAMS,plainnat]{mn2e}

\usepackage{graphicx}
\usepackage{lscape}

\def\Msun{\ifmmode{~{\rm M}_\odot}\else${\rm M}_\odot$~\fi}
\def\kms{\ifmmode{$~km\thinspace s$^{-1}}\else km\thinspace s$^{-1}$\fi}

\def\ee{\end{equation}}
\def\be{\begin{equation}}

\def\kms{km s$^{-1}$}

\def\h2{H$_2$}

\title[Stellar populations in spiral galaxies]{Near-infrared spectroscopy of stellar 
populations in nearby spiral galaxies\thanks{Based on observations collected at the European Southern Observatory (ESO), Chile, programs 068.B-0653 and 076.B-0714}}

\author[J.K. Kotilainen {et~al.}]{J.K. Kotilainen$^{1}$\thanks{E-mail:
jarkot@utu.fi}, T. Hyv\"onen$^{1,2}$, J. Reunanen$^{1,3}$, and V.D. Ivanov$^{4}$ 
\\
$^{1}$ Finnish Centre for Astronomy with ESO (FINCA), University of Turku, V\"ais\"al\"antie 20, FI-21500 Piikki\"o, Finland\\
$^{2}$ Department of Materials Science, Tampere University of Technology, P.O. Box 589, FI-33101 Tampere, Finland\\
$^{3}$ Tuorla Observatory, Department of Physics and Astronomy, University of Turku, V\"ais\"al\"antie 20, FI-21500 Piikki\"o, Finland\\
$^{4}$ European Southern Observatory, Ave. Alonso de Cordova 3107, Vitacura, Santiago 19001, Chile\\
}

\begin{document}

\date{Accepted 2012 May 30. Received 2012 May 30; in original form 2011 August 16}

\pagerange{\pageref{firstpage}--\pageref{lastpage}} \pubyear{2012}
\maketitle
\label{firstpage}

\begin{abstract}
We present high spatial resolution, medium spectral resolution near-infrared 
(NIR) $H$- and $K$-band long-slit spectroscopy for a sample of 29 nearby 
($z<0.01$) inactive spiral galaxies, to study the composition of their NIR 
stellar populations. These spectra contain a wealth of diagnostic stellar 
absorption lines, e.g. Mg{\sc i} 1.575\,$\mu$m, Si{\sc i} 1.588\,$\mu$m, 
CO (6-3) 1.619\,$\mu$m, Mg{\sc i} 1.711\,$\mu$m, Na{\sc i} 2.207\,$\mu$m, 
Ca{\sc i} 2.263\,$\mu$m and the $^{12}$CO and $^{13}$CO bandheads longward of 
2.29\,$\mu$m. We use NIR absorption features to study the stellar population 
and star formation properties of the spiral galaxies along 
the Hubble sequence, and we produce the first high spatial resolution NIR 
$HK$-band template spectra for low redshift spiral galaxies along 
the Hubble sequence. These templates will find applications in a variety of 
galaxy studies. The strength of the absorption lines depends on the luminosity 
and/or temperature of stars and, therefore, spectral indices can be used to 
trace the stellar population of galaxies. The entire sample testifies that 
the evolved red stars completely dominate the NIR spectra, and that the hot 
young star contribution is virtually nonexistent.
\end{abstract}


\section{Introduction}

Disentangling the stellar populations of galaxies is important to 
understanding their evolution and the enhancement of star formation in 
the Universe, since the integrated spectrum of galaxies is sensitive to 
the mass, age, metallicity, dust and star formation history of their dominant 
stellar populations. Despite this complexity, the observed optical spectra of 
galaxies can be accurately reproduced (e.g., Worthey 1994). 
High-quality observational data is paramount to test and improve 
spectrophotometric models, which are used, in turn, to derive the main 
properties of high redshift galaxies.   
Optical Lick spectral indices have usually been used in stellar population 
studies of local galaxies (Worthey \& Ottaviani 1997), but they have problems 
due to blended features and various population of galaxies. 
Near-infrared (NIR, hereafter) spectroscopy has advantages compared to 
optical, because red giant branch (RGB) stars dominate at $2\,\mu$m and 
the wavelength region contains many diagnostic stellar absorption lines 
(e.g., Mg{\sc i} 1.575\,$\mu$m, S{\sc i} $1.589\mu$m, CO(6-3) 1.619\,$\mu$m, 
Mg{\sc i} 1.711\,$\mu$m, Na{\sc i} $2.207\mu$m, Ca{\sc i} 2.263\,$\mu$m and 
CO(2-0) bandhead $>$2.29\,$\mu$m) which can be used as indicators of stellar 
population in terms of their temperature and/or luminosity. The contributions 
of dwarf, giant and supergiant stars can be characterized by different depths 
of their absorption lines and by different shapes of their NIR continuum 
(e.g. Lan\c{c}on {et~al.} 1999). 

Much of the work at NIR wavelengths has focused on unusual objects with either 
active galactic nuclei (AGN) or very strong star formation activity. 
These surveys include those of luminous and ultraluminous infrared galaxies 
(Goldader {et~al.} 1997; Burston, Ward, \& Davies 2001; 
Dannerbauer {et~al.} 2005), various types of starbursts 
(Engelbracht {et~al.} 
1998; 
Vanzi \& Rieke 1997;
Coziol, Doyon, \& Demers 2001; Reunanen, Tacconi-Garman \& Ivanov 2007),
Seyfert galaxies (Ivanov {et~al.} 2000; 
Reunanen, Kotilainen, \& Prieto 2002,2003; 
Riffel {et~al.} 
2009) LINERs (Larkin {et~al.} 1998; 
Alonso-Herrero {et~al.} 2000), and ellipticals (Silva {et~al.} 2008; 
Cesetti {et~al.} 2009). 
However, relatively little NIR spectroscopy exists for inactive nearby 
spiral galaxies. Such surveys are necessary, though, for understanding their 
star formation histories, and for setting a baseline for the NIR emission 
from inactive spiral galaxies. Without such a baseline, the contribution of 
quiescent stellar populations to NIR emission in more exotic systems like 
ultraluminous infrared galaxies and starbursts will remain unknown. 

The two largest previous NIR spectroscopic studies of inactive spirals were 
performed by Mannucci {et~al.} (2001) and Bendo \& Joseph (2004). 
Mannucci {et~al.} (2001) studied 28 galaxies with a 7 x 53 arcsec aperture and 
R = 400 - 500. Generally, they found a high degree of uniformity for 
the spectra of spirals (earlier types being more homogeneous) but their study 
was hampered by the low spectral resolution which prevented them from 
detecting faint and/or blended lines. More recently, Bendo \& Joseph (2004) 
studied 41 galaxies in the K-band and 20 galaxies in the H-band, with slightly 
higher spectral resolution R = 800--1200. They extracted spectra from 
30$\times$15\,arcsec apertures, and found that in almost all cases evolved 
red stars completely dominate the nuclear stellar populations, while young 
stars were virtually non-existent. We stress, however, that both these studies 
suffer from poor spatial resolution, leading to a possible mix of bulge and 
disk stellar populations. 

To improve this situation, we present in this work NIR spectroscopy of nearby 
inactive spiral galaxies along the Hubble sequence, to study their stellar 
population content and SF histories. Stellar emission and absorption features 
can probe with great detail the age and SF properties of the stellar 
population in the nucleus and as a function of radius. Comparison of these 
properties along the Hubble sequence of spirals will assess any relationship 
of SF with Hubble stage. 

We present NIR spectroscopy for a sample of 29 nearby ($z<0.01$) spirals with 
reasonable spatial resolution ($\sim$1 kpc/arcsec) and medium spectral 
resolution ($R\sim$600) to study the NIR properties and ages of their stellar 
populations. Except for a few individual sources, these are the first 
medium resolution NIR spectra of a sizable sample of low redshift spirals. 
Long-slit spectra of the spirals are extracted across the nucleus to detect 
all the important diagnostic absorption lines, and emission from the galaxies 
can usually be studied to large distance from the nucleus. Our aim is to study 
the stellar population of spirals based on the NIR stellar absorption indices, 
and to inter-compare these indices within the different morphology types. 

This paper is organized as follows: In Section\,\ref{sec:sample} we describe 
the sample selection, observations, data reduction and methods of analysis. 
In Section\,\ref{sec:results}, we present the results and discussion 
concerning the properties of the galaxies, and the composite quiescent spectra, 
and in Section\,\ref{sec:conclusions} we summarize our conclusions based on 
the full sample of 29 spiral galaxies.  Throughout this paper, we adopt 
a concordance cosmology with $H_0$ = 70 {\kms} Mpc$^{-1}$, $\Omega_m$ = 0.3 
and $\Omega_\Lambda$ = 0.7.

\section{Sample selection, observations, data reduction and analysis}\label{sec:sample}

This project started as a followup to our NIR spectroscopy of Seyfert galaxies 
(Reunanen {et~al.} 2002, 2003). 
We created a comparison sample of non-active spiral galaxies that were matched in Hubble type with our sample of Seyferts. 
We selected in each Hubble type from E to Sc a similar number of nearby bright 
galaxies (z $<$ 0.01; m(K) $<$ 11 to keep the exposure time reasonable to reach 
the required S/N) and paid careful attention that the spiral galaxies had no indication of nuclear or peculiar activity, based on their multiwavelength 
properties, especially from optical spectroscopy, radio and X-ray data.
We planned to observe five galaxies for each Hubble type between E and Sc, 
in order to obtain a good estimate of the intrinsic spread in the spectra of 
each class, to remove possible individual peculiar characteristics. 
Although our sample is not complete, we consider it as representative of the class of nearby spiral galaxies, and an improvement over the samples in Mannucci et al. (2001) and Bendo \& Joseph (2004).

The H- and K-band spectroscopic observations of the spirals were carried out in 
February 2002 and March 2006 at the 3.5 m ESO New Technology Telescope (NTT) 
using the 1024$\times$1024 px  SOFI camera and spectrograph 
(Moorwood {et~al.} 1998) with pixel scale 0.288 arcsec pix$^{-1}$. 
The average seeing during the observations was $\sim$0.7 $''$ FWHM. The medium 
resolution ($R\sim$600) long-slit $HK$-band spectra were taken along the major 
axis of the galaxies and across the nucleus, with the red grism and slit width 
1.0 arcsec. 
The wavelength range of the spectra is from 1.5 to 2.5\,$\mu$m and the useful 
slit length $\sim$2$'$. The spectra of each target were taken with pairs using 
ABBA observing cycle of alternating galaxy and sky exposures 
nodding $\sim$ 60 arcsec along the slit, 
and slightly moving the target position in the slit 
between exposures, with total integration time of 48 minutes 
(except 32 minutes for NGC 5612), divided into individual exposures of 60 s. 

A literature search was carried out to collect previous measurements of the 
galaxies in our sample: intrinsic velocity dispersion $\sigma$, and optical 
spectral indices Mg$_2$ and H$\beta$ in the Lick system (e.g., Targer {et~al.} 
1998 and references therein). The velocity dispersion and the Mg$_2$ are taken 
from the HyperLEDA compilation\footnote{\tt http://leda.univ-lyon1.fr/} 
(Paturel {et~al.} 2003), 
which has applied unification of the measurements from different sources (i.e. 
Golev \& Prugniel 1998). An average H$\beta$ was taken if more than one 
measurement was available. The measurements are listed in 
Table\,\ref{obsspirals}, together with other properties of the observed 
sample of galaxies (redshift, K-band magnitude, morphological type, 
Hubble type), and the FWHM seeing during the observations. The literature 
references are given in Table\,\ref{obsspirals_ref}.

\small
\begin{table*}
\begin{center}
\caption{General properties and journal of observations of the spirals.
 References for the literature measurements of $\sigma$, Mg$_2$, and H$\beta$ 
are listed in Table \,\ref{obsspirals_ref}.}
\begin{tabular}{@{ }l@{ }l@{ }r@{ }l@{ }l@{ }r@{ }r@{ }r@{ }r@{ }l@{ }l@{}}
\hline
Galaxy   & z        &m(K)~ & Morphology           & type    &~~T~~ & $\sigma$   & Mg$_2$          & H$\beta$      & ~Date        & Seeing\\
         &          &       &                     &         &      &~~~~km/s    &                 &               & ~dd/mm/yyyy  & $''$  \\
(1)      &(2)       & (3)   & (4)                 & (5)     & (6)  & (7)        & (8)             & (9)           & ~(10)        & (11)  \\
\hline
NGC 1617~~~& 0.003546& 9.2~ & SB(s)a              & .SBS1.. &  1.0 & . . .      & . . .           & . . .         & ~20/02/2002~ & 0.9   \\
NGC 2090 & 0.003072 & 10.3~ & SA(rs)c             & .SAT5.. &  5.0 & . . .      & . . .           & . . .         & ~07/03/2006~ & 0.8   \\
NGC 2196 & 0.007742 & 10.3~ & (R')SA(s)a          & PSAS1.. &  1.0 & . . .      & . . .           & . . .         & ~08/03/2006~ & 0.6   \\
NGC 2380 & 0.005944 &  9.0~ & SAB00:              & .LX.0*. & -1.7 & 187$\pm$25 & 0.273$\pm$0.012 & . . .         & ~07/03/2006~ & 0.8   \\
NGC 2502 & 0.003551 &  9.6~ & SAB00(s)            & .LXS0.. & -2.0 & 142$\pm$3  &~0.262$\pm$0.005 &~1.41$\pm$0.25 & ~08/03/2006~ & 0.6   \\
NGC 2566 & 0.005460 &  9.5~ & (R')SB(rs)ab\,pec:~~& PSBT2P* &  2.5 & . . .      & . . .           & . . .         & ~07/03/2006~ & 0.75  \\
NGC 2613 & 0.005591 &  9.6~ & SA(s)b              & .SAS3.. &  3.0 & 169$\pm$12 & . . .           & . . .         & ~22/02/2002~ & 0.9   \\
NGC 2640 & 0.003506 &  8.5~ & SAB0-               & .LX.-.. & -3.0 & . . .      & . . .           & . . .         & ~08/03/2006~ & 0.6   \\
NGC 2775 & 0.004503 &  9.6~ & SA(r)ab             & .SAR2.. &  2.0 & 176$\pm$7  & 0.263$\pm$0.014 & . . .         & ~22/02/2002~ & 0.8   \\
NGC 2781 & 0.006848 &  9.8~ & SAB0+(r)            & .LXR+.. & -1.0 & 144$\pm$15 & 0.204$\pm$0.004 & 2.04$\pm$0.17 & ~08/03/2006~ & 0.5   \\
NGC 2855 & 0.006328 & 10.1~ & (R)SA(rs)0/a        & RSAT0.. &  0.0 & 231$\pm$25 & . . .           & . . .         & ~10/03/2006~ & 0.7   \\
NGC 2935 & 0.007575 &  9.8~ & (R')SAB(s)b         & PSXS3.. &  3.0 & . . .      & . . .           & . . .         & ~22/02/2002~ & 1.0   \\
NGC 3056 & 0.003246 & 10.1~ & (R)SA0-(s):         & RLAS+*. & -1.0 &  74$\pm$3  & 0.155$\pm$0.018 & . . .         & ~07/03/2006~ & 0.75  \\
NGC 3637 & 0.006158 & 10.1~ & (R)SB(r)0           & RLBR0.. & -2.0 & 123$\pm$14 & 0.243$\pm$0.016 & . . .         & ~08/03/2006~ & 0.45  \\
NGC 3715 & 0.007085 & 10.8~ & (R')SB(rs)bc:       & PSBT4*. &  4.0 & . . .      & . . .           & . . .         & ~07/03/2006~ & 0.8   \\
NGC 3810 & 0.003309 & 10.6~ & SA(rs)c             & .SAT5.. &  5.0 &  63$\pm$8  & 0.124$\pm$0.011 & . . .         & ~10/03/2006~ & 0.7   \\
NGC 3892 & 0.005994 & 10.6~ & SB0+(rs)            & .LBT+.. & -1.0 & 116$\pm$6  & 0.248$\pm$0.017 & . . .         & ~20/02/2002~ & 0.9   \\
NGC 4179 & 0.004190 & 11.3~ & S0 edge-on          & .L..../ & -2.0 & 157$\pm$8  & . . .           & . . .         & ~21/02/2002~ & 0.8   \\
NGC 4546 & 0.003502 &  8.9~ & SB0-(s):            & .LBS-*. & -3.0 & 197$\pm$13 & 0.315$\pm$0.008 & . . .         & ~10/03/2006~ & 0.7   \\
NGC 4751 & 0.006985 &  9.7~ & SA0-:               & .LA.-*. & -3.0 & 349$\pm$10 & 0.339$\pm$0.008 & 1.48$\pm$0.13 & ~10/03/2006~ & 0.7   \\
NGC 4856 & 0.004513 &  9.0~ & SB(s)0/a            & .SBS0.. &  0.0 & 160$\pm$4  & 0.241$\pm$0.010 & . . .         & ~07/03/2006~ & 0.8   \\
NGC 4958 & 0.004853 &  9.0~ & SB0(r)? edge-on     & .LBR./  & -2.0 & 156$\pm$4  & 0.264$\pm$0.004 & 1.43$\pm$0.22 & ~08/03/2006~ & 0.5   \\
NGC 5054 & 0.005807 &  9.8~ & SA(s)bc             & .SAS4.. &  4.0 & . . .      & . . .           & . . .         & ~10/03/2006~ & 0.8   \\
IC  4214 & 0.007705 &  9.6~ & (R')SB(r)ab         & PSBR2.. &  1.5 & 173$\pm$5  & 0.241$\pm$0.006 & 1.56$\pm$0.04 & ~22/02/2002~ & 0.8   \\
NGC 5101 & 0.006231 &  9.2~ & (R)SB(rs)0/a        & RSBT0.. &  0.0 & 203$\pm$17 & 0.297$\pm$0.020 & 0.99$\pm$0.35 & ~07/03/2006~ & 0.8   \\
NGC 5134 & 0.005861 & 10.5~ & SA(s)b?             & .SAS3.  &  3.0 & 119$\pm$10 & 0.221$\pm$0.010 & 2.22$\pm$0.05 & ~07/03/2006~ & 0.8   \\
NGC 5507 & 0.006151 & 10.2~ & SAB(r)00            & .LXR0.. & -2.3 & 182$\pm$9  & 0.342$\pm$0.017 & . . .         & ~20/02/2002~ & 0.9   \\
NGC 5612 & 0.009003 &  9.8~ & SAab:               & .SA.2*. &  2.0 & . . .      & . . .           & . . .         & ~08/03/2006~ & 0.7   \\
NGC 5806 & 0.004533 & 10.4~ & SAB(s)b             & .SXS3.. &  3.0 & . . .      & . . .           & . . .         & ~08/03/2006~ & 0.5   \\
\hline 
\end{tabular}
\label{obsspirals}
\end{center}
\begin{list}{}{}
\item[] 
Columns: 
(1) target name; 
(2) redshift from NED; 
(3) the $Ks$-band magnitude from 2MASS in a 10 arcsec diameter aperture; 
(4) and (5) morphology and type from NED and RC3, respectively; 
(6) Hubble type from RC3; 
(7) velocity dispersion; 
(8) and (9) Lick Mg$_2$ and H$\beta$ index, respectively;
(10) observing date; 
and (11) average seeing during the observations.
\end{list}
\end{table*}
\normalsize

\small
\begin{table}
\begin{center}
\caption{References for the central velocity dispersion, Mg$_2$ and H$\beta$ 
measurements from the literature. Only galaxies from our sample with 
available measurements are listed. For convenience, the reference codes are 
identical with the ones used in HyperLEDA (Paturel {et~al.} 2003).}
\begin{tabular}{@{ }l@{ }l@{ }l@{ }l@{ }}
\hline
Galaxy     & \multicolumn{3}{c}{References for:}              \\
           & $\sigma$        & Mg$_2$         & H$\beta$      \\
(1)        & (2)             & (3)            & (4)           \\
\hline
NGC\,2380~~& 7Sam,\,LCO-LO   & 7Sam,\,LCO-LO~ & . . .         \\
NGC\,2502  & 51057,\,Be02,   & 51057,\,CDB-93 & Be02,         \\                                    
           & CDB-93,\,CPWSM~~&                & CDB-93        \\                                    
NGC\,2613  & Bot89b,\,BRBH93 & . . .          & . . .         \\
NGC\,2775  & S83,\,Ton84c,   & JMA96a         & . . .         \\
           & WK,\,WM         &                &               \\
NGC\,2781  & OFJSB           & Tr98a,\,WFG92a & Tr98a         \\
NGC\,2855  & WM              & . . .          & . . .         \\
NGC\,3056  & 51057           & 51057          & . . .         \\
NGC\,3637  & 51057           & 51057          & . . .         \\
NGC\,3810  & HSMP99,\,Vega01 & PMS01b         & . . .         \\
NGC\,3892  & 51057           & 51057          & . . .         \\
NGC\,4179  & DS83,\,ENEARc7  & . . .          & . . .         \\
NGC\,4546  & 51057,\,BGO91   & 51057          & . . .         \\
NGC\,4751  & 51057,\,Be02    & 51057          & Be02          \\
NGC\,4856  & 51057           & 51057          & . . .         \\
NGC\,4958  & 51057,\,DBD93,  & 51057,\,DBD93, & Tr98a         \\
           & OFJSB           & Tr98a          &               \\
IC\,4214   & 51057,{\sc i}FC96   & 51057,{\sc i}FC96  & IFC96         \\
NGC\,5101  & IFC96,\,OFJSB   & IFC96,\,Tr98a  & IFC96,\,Tr98a \\
NGC\,5134  & IFC96           & IFC96          & IFC96         \\
NGC\,5507  & 51057           & 51057          & . . .         \\
\hline 
\end{tabular}
\label{obsspirals_ref}
\end{center}
\begin{list}{}{}
\item[] 
References: 
51057   - Wegner {et~al.} (2003);
7Sam    - Faber {et~al.} (1989);
Be02    - Beuing {et~al.} (2002);
BGO91   - Bettoni {et~al.} (1991);
CDB-93  - Carollo {et~al.} (1993);
DBD93   - De Souza {et~al.} (1993);
DS83    - Dressler \& Sandage (1983);
ENEARc7 - Bernardi {et~al.} (2002);
HSMP99  - H\'eraudeau {et~al.} (1999);
IFC96   - Idiart {et~al.} (1996);
JMA96a  - Jablonka {et~al.} (1996);
LCO-LO  - Davies {et~al.} (1987);
OFJSB   - Dalle Ore {et~al.} (1991);
S83     - Schechter (1983);
Ton84c  - Tonry (1984);
Tr98a   - Trager {et~al.} (1998);
Vega01  - Vega Beltr\'an {et~al.} (2001);
WFG92a  - Worthey {et~al.} (1992);
WK      - Whitmore \& Kirshner (1981);
WM      - Whitmore \& Malumuth (1984).
\end{list}
\end{table}
\normalsize  

Data reduction was performed using IRAF\footnote{IRAF is distributed by the 
National Optical Astronomy Observatories, which are operated by the Association 
of Universities for Research in Astronomy, Inc., under cooperative agreement 
with the National Science Foundation.} and followed the procedure described in 
Hyv\"onen {et~al.} (2009). 
The AB pairs of spectra were subtracted from each other to eliminate background 
sky emission and then divided by the flat-field images, constructed from 
incandescent lamp spectra. Bad pixels and cosmic rays were masked out. 
One-dimensional spectra were extracted from each individual image. 
Wavelength calibration was done using Xe-arc lamp calibration frames
using the same continuum tracing as for the galaxy,
and its zero point was checked against the OH sky lines.  
The galaxy spectra were divided by a G-type 
spectroscopic standard star and then averaged. The spectroscopic standards were 
selected to be as close to the target galaxies as possible. The nuclear 
('nuc', 1\,arcsec diameter, corresponding to $\sim$1 kpc at the average 
distance of the galaxies) and off-nuclear extended emission spectra 
('off1' and 'off2', 
from radii 0.5 to 2 arcsec at opposite sides of the nucleus, and covering an area of 1.5 arcsec$^2$) 
were extracted for each target. 
Three or even more bin widths were added together to trace 
the extended emission. The $H$-band spectra were normalized by fitting the 
continuum level to both sides of the absorption features. In the $K$-band, the 
continuum was fitted only to the shortward wavelengths from the CO bandhead. 
The journal of observations of the sample is presented in 
Table~\ref{obsspirals}.

The equivalent widths (EW) of the spectral lines were measured on the 
normalized spectra, integrating over the same passbands for all targets:
$\lambda$=1.5870-1.5910\,$\mu$m for Si{\sc i}, 
$\lambda$=1.6175-1.6220\,$\mu$m for CO(6-3), 
$\lambda$=2.2040-2.2107\,$\mu$m for Na{\sc i}, 
$\lambda$=2.2577-2.2692\,$\mu$m for Ca{\sc i}, 
and $\lambda$=2.2924-2.2977\,$\mu$m for CO(2-0). 
The continuum at the line was defined by
a straight line fitted to the continuum around the spectral feature. For 
purpose of a direct comparison, the same wavelength ranges were adopted as 
in Origlia {et~al.} (1993) and Silva {et~al.} (2008).

The accuracy of the EWs is limited by the uncertain continuum level in 
the vicinity of the particular line rather than the S/N of the spectra.
This is especially true for the 
$H$-band which is rich in spectral features, and for the CO(2-0) bandhead in 
the $K$-band because the continuum level has to be extrapolated longward of 
2.295\,$\mu$m. It is worth to keep in mind, that there are many blends 
and broad molecular features (Wallace \& Hinkle 1996). 
Furthermore, the measured EWs of the indices are evidently upper limits 
due to spreading caused by velocity dispersion. The possible dependence of the 
EWs on velocity dispersion and resolution was checked by degrading the medium 
resolution stellar spectra from Wallace \& Hinkle (1997) to the lower 
resolution applicable for the spiral galaxy sample, and measuring the indices 
of all the five diagnostic features for both resolutions. This comparison 
showed that the indices do not significantly depend on the degradation: 
all measurements agree within the errors. Therefore, the bandpasses are 
wide enough to fully encompass the measured features, despite 
the broadening due to the intrinsic velocity dispersion. 
We estimate our typical EW errors to be: $\sim$0.5 $\AA$ for the $H$-band and 
for the $K$-band metallic lines, and $\sim$1.0 $\AA$ for the $K$-band 
CO(2-0) feature.

\section{Results and discussion}\label{sec:results}

The inactive spiral galaxies have strikingly similar NIR nuclear spectra as a 
function of their Hubble type, especially when compared to more active classes 
of objects. The main $H$- and $K$-band spectral absorption features from the 
stellar population -- Mg{\sc i}, Si{\sc i}, CO(6-3), Na{\sc i}, Ca{\sc i}, and 
CO(2-0) -- are clearly 
visible. Their relative strengths are quite similar for all Hubble types.
No H-band [Fe II] or K-band Br$\gamma$ or H$_2$ emission lines are evident 
at a significant level. However, weak H$_2$ and/or Br$\gamma$ emission was 
detectable in a few galaxies after subtracting an elliptical galaxy 
template representing the old stellar population.
(see Section\,\ref{sec:emission}). 

\begin{figure*}
\begin{center}
\includegraphics[width=0.90\textwidth]{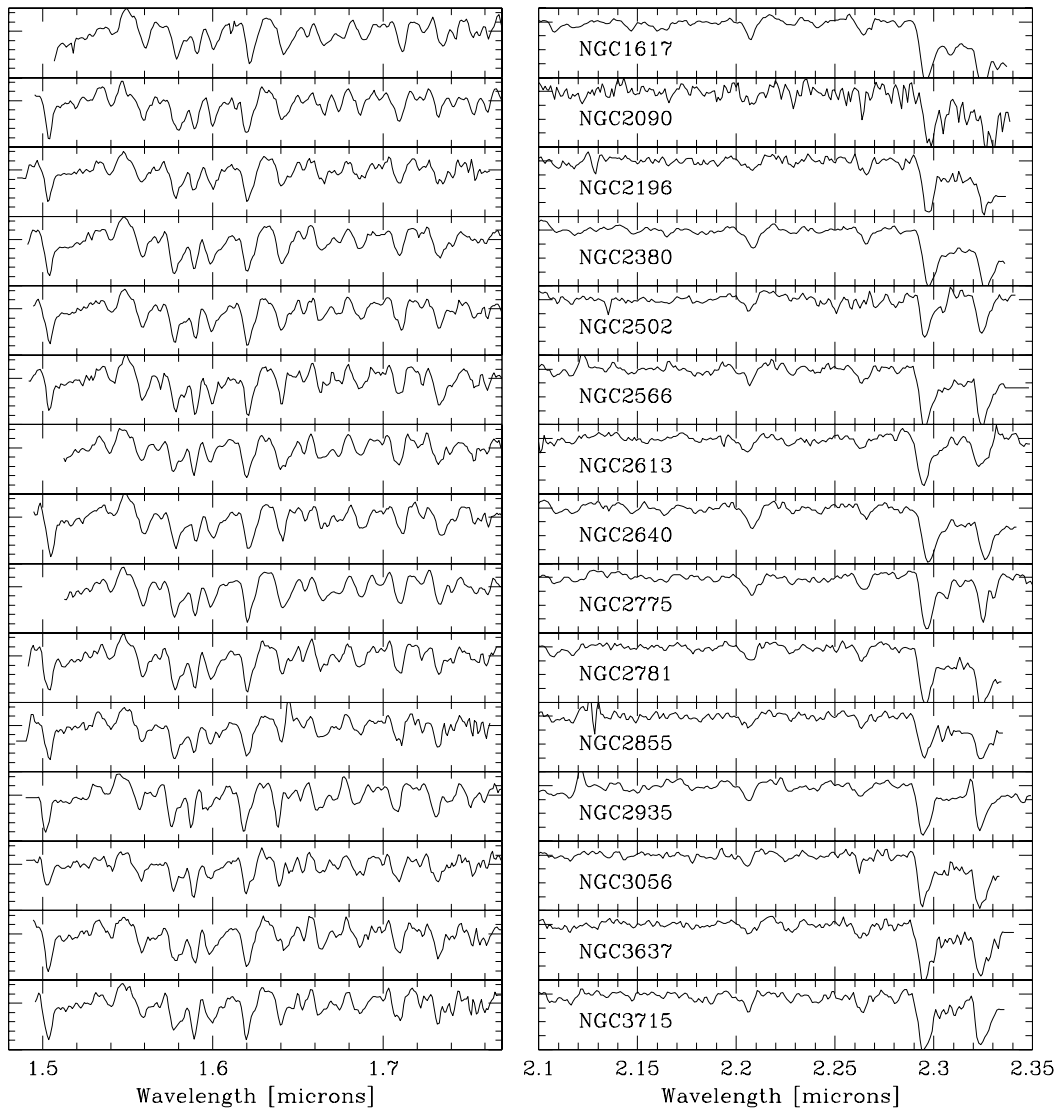}
\caption{The $HK$-band nuclear spectra of the individual spiral galaxies. 
}
\label{spectra}
\end{center}
\end{figure*}
\addtocounter{figure}{-1}

\begin{figure*}
\begin{center}
\includegraphics[width=0.90\textwidth]{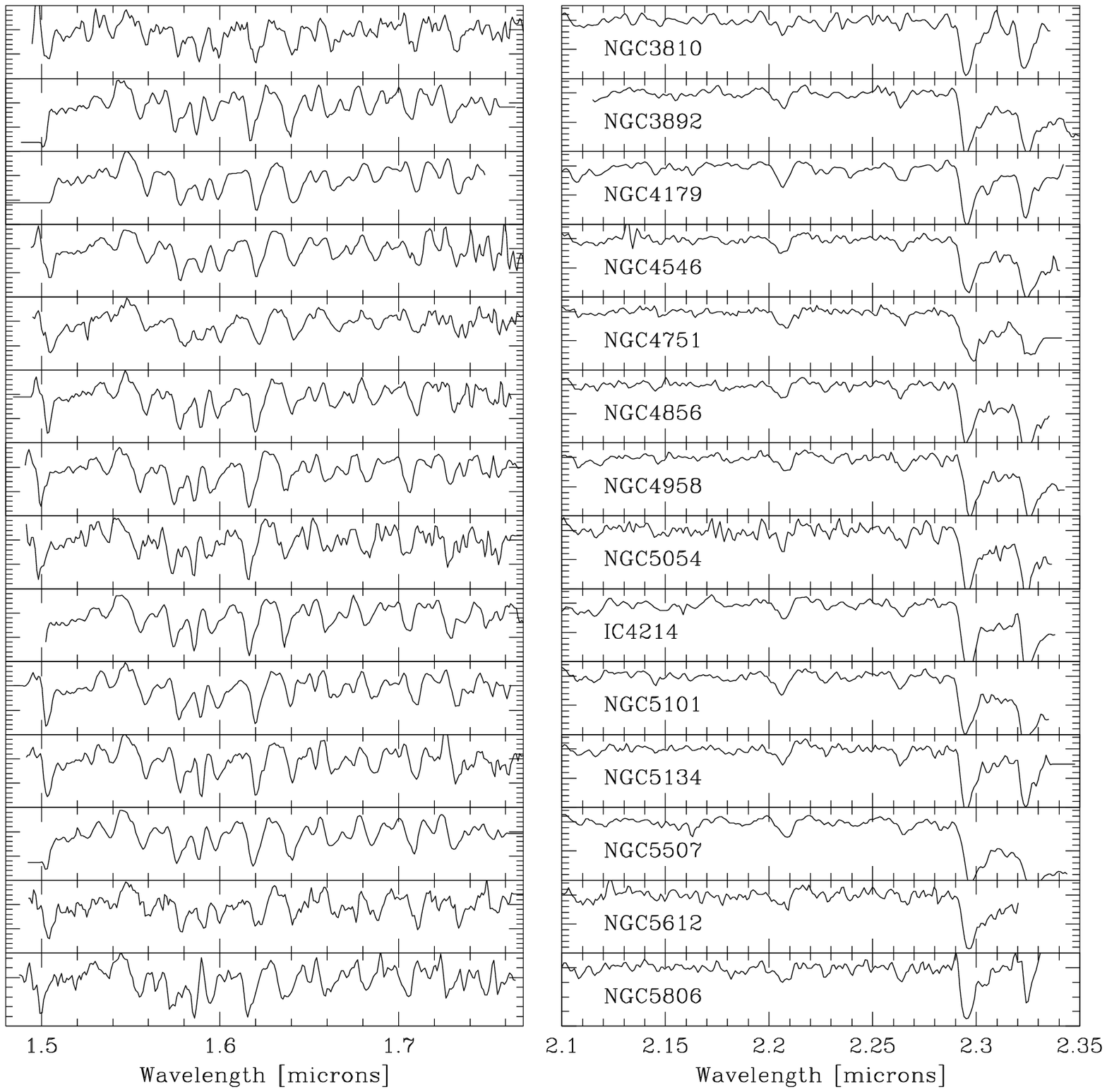}
\caption{
continued.
The $HK$-band nuclear spectra of the individual spiral galaxies. 
}
 \label{spectra}
\end{center}
\end{figure*}

The $HK$-band nuclear spectra for the individual spiral galaxies in 
the sample are presented in Fig.~\ref{spectra}.
The EWs of all the diagnostic absorption lines were measured for all 
the spiral galaxies, and they are presented in Tables~\ref{ewh} and \ref{ewk}. 
The off-nuclear and nuclear EWs of the spirals are generally similar, 
and they are in broad agreement with the EW values from the literature (Mannucci {et~al.} 2001; Bendo \& Joseph 2004; Silva {et~al.} 2008; Cesetti {et~al.} 2009; Hyv\"onen {et~al.} 2009), if differences in aperture sizes and index definitions are taken into account. 

\begin{table*}
\begin{center}
\caption{EWs of the H-band absorption lines for the nuclear and off-nuclear
 spectra of the spiral galaxies in units of $\AA$.}
\begin{tabular}{llllllllllllll}
\hline
\noalign{\smallskip}
Galaxy & T & \multicolumn{3}{c}{Mg{\sc i}} & \multicolumn{3}{c}{Si{\sc i}} & \multicolumn{3}{c}{CO(6-3)} & \multicolumn{3}{c}{Mg{\sc i}} \\
$\lambda$ [$\mu$m] & T & \multicolumn{3}{c}{1.575} & \multicolumn{3}{c}{1.589} & \multicolumn{3}{c}{1.619} & \multicolumn{3}{c}{1.711} \\
       &   & nuc & off1 & off2& nuc & off1 & off2 & nuc & off1 & off2 & nuc & off1 & off2 \\
(1) & (2) & (3) & (4) & (5) & (6) & (7) & (8) & (9) & (10) & (11) & (12) & (13) & (14) \\
\noalign{\smallskip}
\hline
\noalign{\smallskip}
\multicolumn{14}{l}{Sc spirals (4 $\leq$ T $<$ 5)}\\
NGC 2090 &  5.0 & 3.7 & 3.5 & 3.4 & 3.5 & 3.8 & 4.2 & 5.5 & 5.0 & 6.0 & 3.9 & 3.1 & 3.5 \\
NGC 3715 &  4.0 & 3.5 & 3.4 & 3.3 & 4.0 & 3.3 & 3.3 & 4.8 & 5.3 & 4.6 & 3.0 & 3.3 & 3.2 \\
NGC 3810 &  5.0 & 3.4 & 3.0 & 3.3 & 3.5 & 3.8 & 3.8 & 5.3 & 5.0 & 5.2 & 3.3 & 3.3 & 3.2 \\
NGC 5054 &  4.0 & 2.7 & 2.9 & 2.7 & 3.1 & 3.2 & 3.5 & 5.4 & 6.6 & 6.6 & 4.0 & 3.8 & 3.1 \\
\multicolumn{14}{l}{}\\
\multicolumn{14}{l}{Sb spirals (2.0 $\leq$ T $<$ 4)}\\
NGC 2566 &  2.5 & 2.4 & 2.7 & 1.8 & 3.3 & 3.6 & 3.6 & 3.6 & 4.8 & 5.4 & 4.2 & 3.3 & 3.6 \\
NGC 2613 &  3.0 & ... & ... & ... & 3.1 & 3.2 & 3.2 & 4.9 & 4.6 & 4.8 & 4.5 & 3.9 & 3.5 \\
NGC 2775 &  2.0 & ... & ... & ... & 3.8 & 3.6 & 3.8 & 5.0 & 4.9 & 4.8 & 4.1 & 3.8 & 3.9 \\
NGC 2935 &  3.0 & 3.1 & 3.0 & 2.8 & 3.4 & 3.0 & 3.6 & 5.2 & 5.7 & 5.4 & 3.6 & 3.3 & 3.2 \\
NGC 5134 &  3.0 & 3.2 & 3.4 & 3.3 & 3.3 & 3.2 & 3.2 & 4.8 & 5.0 & 5.0 & 3.8 & 3.5 & 3.3 \\
NGC 5612 &  2.0 & 3.2 & 3.3 & 3.3 & 2.7 & 2.5 & 2.1 & 4.5 & 5.0 & 4.6 & 4.2 & 3.6 & 3.6 \\
NGC 5806 &  3.0 & 3.1 & 3.6 & 3.7 & 3.3 & 3.5 & 3.3 & 5.5 & 5.2 & 5.6 & 3.7 & 3.5 & 3.4 \\
\multicolumn{14}{l}{}\\
\multicolumn{14}{l}{Sa spirals (0.0 $\leq$ T $<$ 2.0)}\\
NGC 1617 &  1.0 & ... & ... & ... & 3.5 & 3.4 & 3.4 & 5.2 & 5.3 & 5.1 & 3.4 & 3.4 & 3.4 \\
NGC 2196 &  1.0 & 3.0 & 3.1 & 3.3 & 3.5 & 2.8 & 3.0 & 5.2 & 5.5 & 4.6 & 2.9 & 2.6 & 2.5 \\
NGC 2855 &  0.0 & 3.2 & 3.6 & 3.6 & 2.6 & 3.4 & 3.5 & 4.3 & 4.4 & 4.6 & 2.9 & 3.1 & 3.2 \\
NGC 4856 &  0.0 & 3.1 & 3.1 & 3.1 & 3.4 & 3.5 & 3.4 & 5.0 & 4.8 & 5.1 & 3.8 & 3.8 & 3.5 \\
IC 4214  &  1.5 & 3.4 & 3.2 & 3.0 & 4.0 & 3.2 & 3.4 & 4.2 & 4.2 & 4.7 & 4.1 & 3.8 & 3.6 \\
NGC 5101 &  0.0 & 3.3 & 3.4 & 3.5 & 3.0 & 2.8 & 2.6 & 5.4 & 5.0 & 5.1 & 4.1 & 3.4 & 3.4 \\
\multicolumn{14}{l}{}\\
\multicolumn{14}{l}{Late S0 (-1.7 $\leq$ T $<$ 0.0)}\\
NGC 2380 & -1.7 & 4.1 & 3.8 & 3.8 & 3.6 & 3.2 & 3.2 & 4.9 & 4.6 & 5.0 & 3.4 & 3.2 & 3.1 \\
NGC 2781 & -1.0 & 3.8 & 3.6 & 3.6 & 3.9 & 3.4 & 3.5 & 5.0 & 4.4 & 4.8 & 3.3 & 3.5 & 3.6 \\
NGC 3056 & -1.0 & 2.4 & 2.7 & 2.4 & 3.3 & 3.2 & 3.2 & 4.9 & 4.6 & 4.6 & 3.4 & 3.4 & 3.4 \\
NGC 3892 & -1.0 & ... & ... & ... & 3.6 & 3.6 & 3.7 & 5.5 & 5.4 & 5.5 & 3.7 & 3.5 & 3.4 \\
\multicolumn{14}{l}{}\\
\multicolumn{14}{l}{Intermediate S0 (-2.0 $\leq$ T $<$ -1.7)}\\
NGC 2502 & -2.0 & 3.0 & 3.0 & 3.0 & 3.3 & 3.1 & 2.9 & 5.4 & 4.8 & 4.9 & 3.2 & 3.8 & 3.3 \\
NGC 3637 & -2.0 & 3.3 & 3.3 & 3.4 & 3.5 & 3.8 & 4.0 & 5.0 & 4.8 & 5.4 & 4.0 & 3.1 & 3.2 \\
NGC 4179 & -2.0 & ... & ... & ... & 3.1 & 3.0 & 3.1 & 5.1 & 4.8 & 4.8 & 3.4 & 3.4 & 3.3 \\
NGC 4958 & -2.0 & 2.6 & 2.6 & 2.5 & 3.3 & 3.2 & 3.4 & 5.4 & 5.1 & 5.4 & 3.7 & 3.7 & 3.7 \\
\multicolumn{14}{l}{}\\
\multicolumn{14}{l}{Early S0 (-3.0 $\leq$ T $<$ -2.0)}\\
NGC 2640 & -3.0 & 3.9 & 3.6 & 3.0 & 3.1 & 3.4 & 2.8 & 5.0 & 4.8 & 4.2 & 3.4 & 3.4 & 3.1 \\
NGC 4546 & -3.0 & 3.1 & 3.0 & 3.2 & 3.3 & 3.0 & 3.1 & 4.8 & 4.5 & 4.6 & 3.3 & 2.9 & 2.9 \\
NGC 4751 & -3.0 & 3.5 & 3.1 & 3.4 & 2.8 & 2.9 & 2.7 & 4.3 & 4.5 & 4.8 & 3.6 & 3.3 & 3.0 \\
NGC 5507 & -3.0 & ... & ... & ... & 2.9 & 3.1 & 3.2 & 6.1 & 6.1 & 5.9 & 3.7 & 4.0 & 4.1 \\
\noalign{\smallskip}
\hline
\end{tabular}
\label{ewh}
\end{center}
\begin{list}{}{}
\item[] 
Column (1) gives the name of the spiral galaxy; (2) the luminosity class T; 
and (3) to (14) the EWs of the main diagnostic absorption lines in 
units of $\AA$, in the nuclear 1.0 arcsec diameter aperture (nuc), and in 
the two off-nuclear apertures (off1 and off2). 
\end{list}
\end{table*}

\begin{table*}
\begin{center}
\caption{EWs of the K-band absorption and emission lines for the nuclear and 
off-nuclear spectra of the spiral galaxies in units of $\AA$.}
\begin{tabular}{llllllllllllll}
\hline
\noalign{\smallskip}
Galaxy & T & \multicolumn{3}{c}{Br$\gamma$} & \multicolumn{3}{c}{Na{\sc i}} & \multicolumn{3}{c}{Ca{\sc i}} & \multicolumn{3}{c}{CO(2-0)} \\
$\lambda$ [$\mu$m] & T & \multicolumn{3}{c}{2.166} & \multicolumn{3}{c}{2.207} & \multicolumn{3}{c}{2.263} & \multicolumn{3}{c}{$>$2.295} \\
       &   & nuc & off1 & off2 & nuc & off1 & off2 & nuc & off1 & off2 & nuc & off1 & off2 \\
(1) & (2) & (3) & (4) & (5) & (6) & (7) & (8) & (9) & (10) & (11) & (12) & (13) & (14) \\
\noalign{\smallskip}
\hline
\noalign{\smallskip}
\multicolumn{14}{l}{Sc spirals (4 $\leq$ T $<$ 5)}\\
NGC 2090 &  5.0 & ... & ... & ... & 2.0 & 2.6 & 3.0 & 3.1 & 2.7 & 2.6 & 14.9 & 15.4 & 14.8 \\
NGC 3715 &  4.0 & ... & 0.6 & ... & 3.1 & 3.4 & 3.2 & 3.7 & 3.7 & 3.7 & 15.3 & 15.8 & 17.4 \\
NGC 3810 &  5.0 & ... & ... & ... & 2.6 & 2.4 & 3.0 & 1.7 & 2.7 & 3.5 & 15.9 & 13.7 & 14.5 \\
NGC 5054 &  4.0 & ... & ... & ... & 3.9 & 3.7 & 3.8 & 3.6 & 3.4 & 2.7 & 15.2 & 15.5 & 16.1 \\
\multicolumn{14}{l}{}\\
\multicolumn{14}{l}{Sb spirals (2.0 $\leq$ T $<$ 4)}\\
NGC 2566 &  2.5 & 0.6 & 5.4 & 7.4 & 3.8 & 3.2 & 3.5 & 2.6 & 2.2 & 2.6 & 15.0 & 15.4 & 14.8 \\
NGC 2613 &  3.0 & ... & ... & ... & 3.8 & 3.6 & 2.6 & 2.0 & 2.8 & 2.4 & 15.9 & 15.8 & 15.8 \\
NGC 2775 &  2.0 & ... & ... & 0.3 & 4.5 & 4.0 & 3.5 & 3.5 & 3.0 & 2.6 & 13.8 & 13.0 & 13.2 \\
NGC 2935 &  3.0 & ... & 0.5 & 0.5 & 4.0 & 3.1 & 3.5 & 2.8 & 2.6 & 3.2 & 15.1 & 16.2 & 15.5 \\
NGC 5134 &  3.0 & ... & ... & ... & 3.2 & 3.2 & 3.2 & 2.6 & 3.9 & 3.2 & 14.8 & 16.2 & 16.2 \\
NGC 5612 &  2.0 & ... & ... & ... & 4.3 & 3.0 & 3.4 & 2.3 & 2.0 & 2.3 & 15.3 & 14.8 & 16.4 \\
NGC 5806 &  3.0 & ... & ... & 0.6 & 4.1 & 4.2 & 4.2 & 2.7 & 2.6 & 2.8 & 14.1 & 14.0 & 14.6 \\
\multicolumn{14}{l}{}\\
\multicolumn{14}{l}{Sa spirals (0.0 $\leq$ T $<$ 2.0)}\\
NGC 1617 &  1.0 & ... & ... & ... & 3.0 & 2.9 & 3.3 & 2.8 & 2.6 & 2.8 & 15.7 & 15.8 & 13.8 \\
NGC 2196 &  1.0 & ... & 0.4 & ... & 3.3 & 3.2 & 2.5 & 3.8 & 3.4 & 3.2 & 15.8 & 14.2 & 15.5 \\
NGC 2855 &  0.0 & ... & ... & ... & 2.5 & 3.8 & 3.2 & 2.7 & 2.8 & 2.4 & 12.5 & 14.0 & 14.2 \\
NGC 4856 &  0.0 & ... & ... & ... & 3.5 & 3.3 & 3.6 & 2.6 & 3.1 & 2.2 & 16.9 & 16.6 & 15.2 \\
IC 4214  &  1.5 & 0.8 & 0.7 & 0.4 & 4.4 & 4.0 & 4.0 & 2.5 & 3.0 & 2.8 & 17.8 & 17.5 & 17.7 \\
NGC 5101 &  0.0 & ... & ... & ... & 5.5 & 4.8 & 4.0 & 2.7 & 3.1 & 2.3 & 15.7 & 16.2 & 16.2 \\
\multicolumn{14}{l}{}\\
\multicolumn{14}{l}{Late S0 (-1.7 $\leq$ T $<$ 0.0)}\\
NGC 2380 & -1.7 & ... & ... & ... & 5.2 & 5.4 & 4.0 & 2.9 & 3.0 & 3.1 & 16.3 & 15.5 & 16.6 \\
NGC 2781 & -1.0 & ... & ... & ... & 4.8 & 4.2 & 2.6 & 2.6 & 3.6 & 3.6 & 14.6 & 16.4 & 16.5 \\
NGC 3056 & -1.0 & ... & ... & ... & 2.4 & 2.8 & 2.5 & 1.6 & 2.8 & 3.4 & 13.0 & 13.2 & 15.3 \\
NGC 3892 & -1.0 & 0.3 & 0.7 & 3.2 & 3.4 & 3.2 & 3.0 & 2.8 & 3.0 & 2.6 & 15.1 & 14.7 & 13.6 \\
\multicolumn{14}{l}{}\\
\multicolumn{14}{l}{Intermediate S0 (-2.0 $\leq$ T $<$ -1.7)}\\
NGC 2502 & -2.0 & ... & ... & ... & 4.2 & 3.4 & 3.4 & ... & 2.0 & 2.6 & 9.2 & 15.0 & 14.5 \\
NGC 3637 & -2.0 & ... & ... & ... & 3.5 & 3.2 & 3.5 & 3.2 & 2.0 & 2.4 & 15.1 & 15.8 & 15.4 \\
NGC 4179 & -2.0 & ... & ... & ... & 5.5 & 5.1 & 3.6 & 3.8 & 3.8 & 3.2 & 15.5 & 15.0 & 14.9 \\
NGC 4958 & -2.0 & ... & ... & ... & 4.2 & 4.0 & 3.2 & 3.5 & 3.3 & 2.9 & 13.9 & 14.8 & 15.0 \\
\multicolumn{14}{l}{}\\
\multicolumn{14}{l}{Early S0 (-3.0 $\leq$ T $<$ -2.0)}\\
NGC 2640 & -3.0 & ... & ... & ... & 5.1 & 5.2 & 4.8 & 2.4 & 2.6 & 2.2 & 14.7 & 13.9 & 14.8 \\
NGC 4546 & -3.0 & ... & ... & ... & 3.9 & 3.6 & 4.0 & 2.1 & 2.7 & 2.9 & 15.9 & 15.8 & 16.8 \\
NGC 4751 & -3.0 & ... & ... & ... & 4.9 & 4.7 & 4.1 & 2.1 & 2.5 & 2.6 & 17.6 & 16.4 & 16.9 \\
NGC 5507 & -3.0 & ... & ... & ... & 4.2 & 4.0 & 2.9 & 3.1 & 2.9 & 2.8 & 17.6 & 17.2 & 16.8 \\
\noalign{\smallskip}
\hline
\end{tabular}
\label{ewk}
\end{center}
\begin{list}{}{}
\item[] 
Column (1) gives the name of the spiral galaxy; (2) the luminosity class T; 
and (3) to (23) the EWs of the main diagnostic emission and absorption lines in 
units of $\AA$, in the nuclear 1.0 arcsec diameter aperture (nuc), and in 
the two off-nuclear apertures (off1 and off2). 
\end{list}
\end{table*}

\subsection{NIR line indices as stellar population indicators}

The NIR spectroscopy is well-suited to study stellar populations in 
galaxies and star clusters because it 
contains many strong stellar absorption features tracing the presence of red 
stars, particularly red supergiants and giants. The $K$-band has been widely 
studied (e.g., Merrill \& Ridgway 1979; Kleinmann \& Hall 1986; Wallace \& 
Hinkle 1997, Ivanov {et~al.} 2000), and it is a natural choice for studying 
reddened (dusty) galaxies at low redshift (e.g., Mobasher \& James 2000; 
Mannucci {et~al.} 2001). However, the dust emission that tends to dilute the 
stellar absorption has even smaller effect in the $H$-band which has been 
studied systematically only recently (e.g., Origlia {et~al.} 1993; 
Oliva {et~al.} 1995; Ivanov {et~al.} 2004; Fremaux {et~al.} 2007). 
The $J$-band remains relatively unexplored in galaxies (but see 
Wallace {et~al.} 2000 for stars).

The integrated NIR light is dominated by different stars from those that 
contribute significantly at visible wavelengths: Whereas the integrated light 
at visible wavelengths comes from a mix of stellar types, the $K$-band light 
in all but the youngest stellar systems is dominated by evolved giants and 
asymptotic giant branch (AGB) stars, whose photometric and spectroscopic 
properties are sensitive to age and metallicity (Maraston 2005).

The relative strengths of the NIR features can be used to characterize the 
stellar contents of the central regions of galaxies, and to provide insights 
into their history (e.g., Vazquez {et~al.} 2003) because of their sensitivity 
to the spectral type (temperature) and/or luminosity (gravity) of the star 
(Kleinmann \& Hall 1986; Ali {et~al.}, 1995; Ramirez {et~al.} 1997; Ivanov 
{et~al.} 2004). Ivanov (2001) discussed the possibility to develop NIR analog
of the optical H$\beta$ vs. Mg$_2$ diagnostic diagram used to derive ages and
metallicities. To give an example for the utility of this diagnostic tool: 
while H$\beta$ vs. Mg$_2$ can be used to constrain the properties of the host 
galaxies of optical supernovae (e.g, Hamuy {et~al.} 2000), NIR diagnostic is 
needed to do the same for the host galaxies of obscured supernovae 
(e.g, Mannucci {et~al.} 2003). Ivanov (2001) pointed that the optical 
H$\beta$ has an IR counterpart -- Br$\gamma$ at 2.167\,$\mu$m, and that there 
are a number of candidates to replace the Mg$_2$, including some NIR 
Mg features.

The most evident absorption features in the $K$-band are the molecular CO(2-0) 
bandhead ($>$2.295\,$\mu$m), atomic Br$\gamma$ (2.166\,$\mu$m), Na{\sc i} 
(2.207\,$\mu$m), and Ca{\sc i} (2.263\,$\mu$m) lines. Br$\gamma$ is 
the strongest 
absorption feature within stars earlier than K5, but disappears in stars 
later than K5 (Kleinmann \& Hall 1986), while Na{\sc i} is the strongest atomic 
feature when effective temperature is T$_{\rm eff}$$<$3400\,K (Ali {et~al.} 
1995). The $H$-band has a very complex absorption line pattern due to a number 
of metallic and molecular lines, especially in cool late-type stars. The most 
studied $H$-band features are atomic Si{\sc i} (1.589\,$\mu$m) and molecular 
CO(6-3) (1.619\,$\mu$m) (e.g., Origlia {et~al.} 1993). The CO is very strong 
in young giant and supergiant stars (10 Myr-100 Myr) and strong in cool AGB 
stars (100 Myr-1 Gyr; Oliva {et~al.} 1995), while it is weaker in 
older population. 
This makes it suitable to trace recent star formation in galaxies (Mayya 1997).
There are also some relatively weak Mg{\sc i} features 
(e.g. 1.504, 1.710 $\mu$m) 
that potentially could provide a direct comparison with the optical $\alpha$ 
element abundances, but they are rather demanding in terms of the quality of 
the observations (Cesetti {et~al.} 2009). The Na{\sc i}, Ca{\sc i} and 
CO features in K and M stars become stronger with lower temperature, 
i.e. redder $J$$-$$K$ colour (Ramirez {et~al.} 1997; F\"orster Schreiber 2000; 
Frogel {et~al.} 2001). At a given $J$$-$$K$, giants in more metal rich 
clusters have stronger Na{\sc i} and 
Ca{\sc i} features (Frogel {et~al.} 2001). Some line ratios, such as 
EW(CO 1.62)/EW(Si{\sc i} 1.59) and EW(CO 1.62)/EW(CO 2.29), are even better 
temperature indicators than single lines (Origlia {et~al.} 1993; 
Ivanov {et~al.} 
2004), especially if the two lines have very similar wavelengths making their 
ratio insensitive to dilution from dust or hot stars. Last but not least. most 
of these features suffer some degree of contamination from other elements, much 
like in the optical (e.g., Worthey {et~al.} 1994).

Currently, there are few self-consistent theoretical spectral synthesis 
models for the interpretation of NIR spectra of galaxies (although see e.g. 
Leitherer {et~al.} 1999; Bendo \& Joseph 2004; Maraston 2005; 
Riffel {et~al.} 2008). 
Thus, one has to rely on e.g. their comparison with high-resolution NIR stellar 
spectral atlases (e.g., Wallace \& Hinkle 1996) or on direct comparison 
with optical indices (e.g. Silva {et~al.} 2008, Cesetti {et~al.} 2009).


\subsection{NIR-to-optical and NIR-to-NIR line index relations of spiral galaxies}


The optical Mg$_2$, Fe, and H$\beta$ indices show tight correlations with 
the central velocity dispersion $\sigma$ (e.g. Bernardi {et~al.} 1998; 
Mehlert {et~al.} 2003), suggesting that the chemical and dynamical evolution 
of galaxies are intertwined. 
For the galaxies in our sample, the relation with $\sigma$ is well established 
for Mg$_2$ but it suffers from somewhat more scatter for the Fe5270. 
None of the NIR features show as clear a relation with $\sigma$ as the Mg$_2$. 
The best Mg$_2$ analog in the NIR appears to be Na{\sc i} at 2.207\,$\mu$m 
which shows approximately the same scatter with $\sigma$ as the Fe5270.
We also find that the Na{\sc i} at 2.207\,$\mu$m feature shows a fair direct 
correlation with Mg$_2$ despite the possible combined production of Na in both 
low and high mass stars, which sets it apart from Mg, which is produced only 
in massive stars.

\subsection{Dilution of the NIR line indices of spiral galaxies}

\begin{figure*}
\begin{center}
\includegraphics[width=0.98\textwidth]{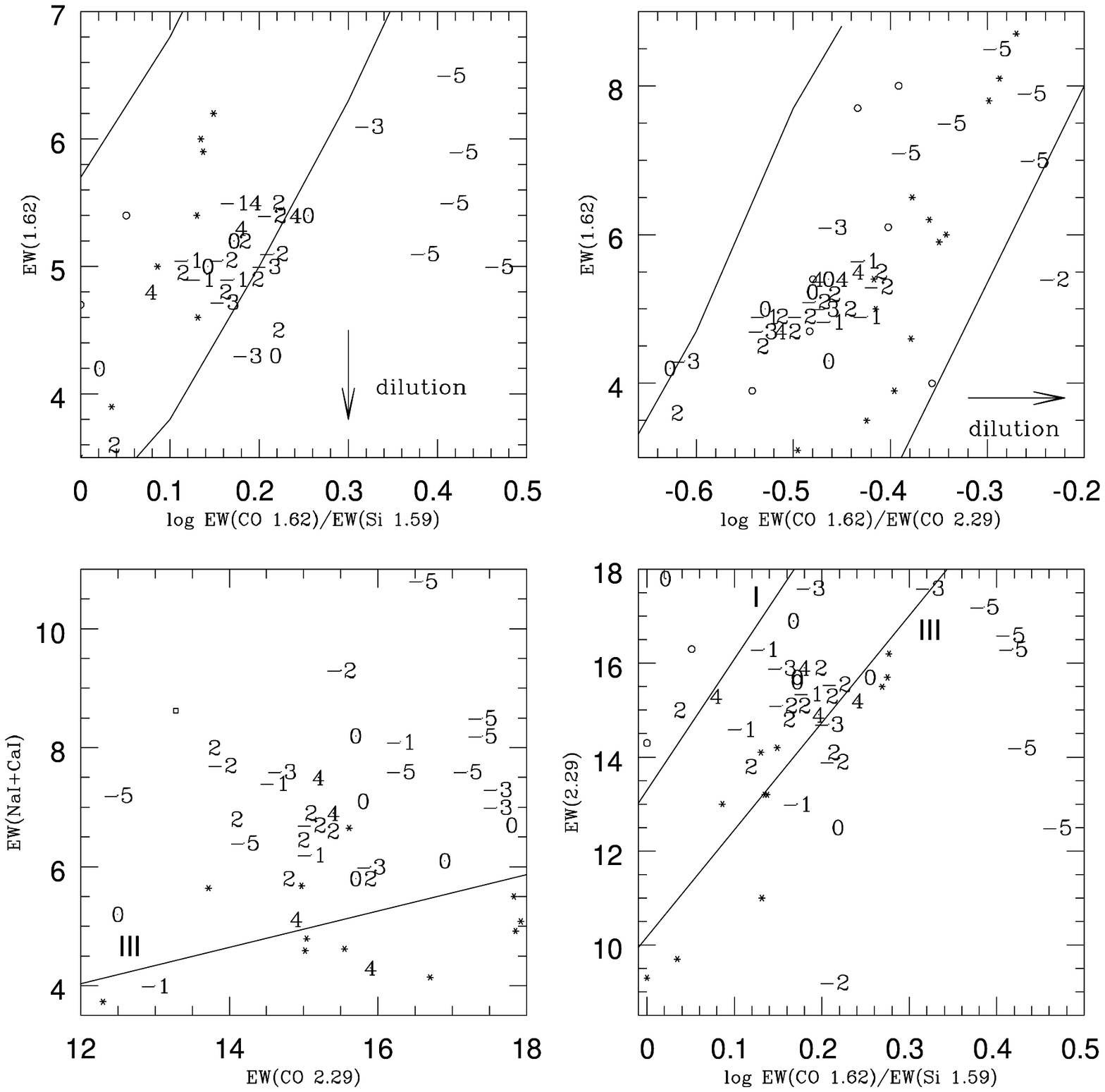}
\caption{For the nuclear spectra of the spiral galaxies (this work) and 
elliptical galaxies (from Hyv\"onen {et~al.} 2009): 
{\bf Top left:} EW(CO 1.62) vs. log[EW(CO 1.62)/(EW(Si 1.59);  
{\bf Top right:} EW(CO 1.62) vs. log[EW(CO 1.62)/(EW(CO 2.29);
{\bf Bottom left:} EW(Na{\sc i} 2.207 + Ca{\sc i} 2.263) vs. EW(CO 2.29); and 
{\bf Bottom right:} EW(CO 2.29) vs. log[EW(CO 1.62)/(EW(Si 1.59).
The individual galaxies are marked according to their Hubble type T 
(from -5 to 4; E to Sc). 
Different luminosity types of stars are shown as small symbols: supergiants I 
(open circles), giants III (asterisks) from Ramirez {et~al.} (1997) 
and dwarfs V (open squares) from Ali {et~al.} (1995). 
The lines in the first two panels enclose the area occupied by stars with no 
dilution and the arrow gives the direction of the effects of dilution. 
In the last two panels, the lines show the relations for different luminosity 
types of stars.
\label{ewt}}
\end{center}
\end{figure*}

Non-stellar thermal sources, such as dust surrounding young star forming 
regions, and/or non-thermal dilution reduce the intrinsic values of EWs 
originating from the stellar population. However, the dust observed in NIR is 
close to its sublimation temperature, and in such cases HII emission line 
series are expected to be observed. The comparison of absorption indices in 
different filters depends on the level of dilution by the non-stellar 
component that can be different in $H$- and $K$-band. The dilution fraction 
of the continuum emission at 1.6\,$\mu$m and 2.3\,$\mu$m can be estimated 
from the plots of EW(CO 1.62) vs. EW(CO 1.62)/EW(Si 1.59), and EW(CO 1.62) 
vs. EW(CO 1.62)/EW(CO 2.29), shown in Fig.~\ref{ewt}, top panels. 
Objects with diluted stellar features lie away from the loci occupied 
by stars, along the dilution vectors,  
and the fraction of non-stellar continuum is simply given by the displacement 
of the point in the diagram (e.g., for dilution from non-thermal emission 
in AGN see Fig.\,3 in Ivanov {et~al.} 2000).
Fig.~\ref{ewt} shows that in the $H$- and $K$-band, 
all spirals have either no or only slight dilution.
This indicates that no significant amount of dust is present in 
the nuclear region of spirals.


Fig.\,\ref{ewt}, bottom left panel, shows the EW(Na{\sc i}+Ca{\sc i}) 
as a function of EW(CO 2.29) for the nuclear spirals, and for 
elliptical galaxies (from Hyv\"onen {et~al.} 2009), together with stars of 
different spectral type. Caution should be exercised when 
comparing stellar and galactic data, as galaxies are composite stellar systems 
and their spectra show the integrated contributions from stars spanning a 
range of properties. Still, given that the NIR spectral region is dominated by 
evolved (RGB) stars, this issue is not as critical as at visible wavelengths, 
where stars contribute over a much larger range of evolutionary states (both 
RGB and main sequence stars (e.g., Maraston 2005). All three luminosity classes 
of stars have a relatively tight linear correlation between 
EW(Na{\sc i}+Ca{\sc i}) 
and EW(CO 2.29). Most of the off-nuclear spirals are in good agreement with the 
relation of giant (III) stars, while ellipticals deviate more strongly from 
that relation. As expected, for spirals in most cases the main role is played 
by class III giant stars, possibly with some contribution from class V dwarf 
stars. 

On the other hand, the plot of EW(CO 2.29) vs. EW(CO 1.62)/EW(Si 1.59) 
(Fig.~\ref{ewt}, bottom right panel) indicates that the stellar populations 
are mostly dominated by a mixture of giants (III) and supergiants (I). 
However, it is very likely that the CO(1.62) line is underestimated and/or 
the Si{\sc i} line overestimated decreasing the line ratio of 
EW(Si{\sc i})/EW(CO 1.62) 
and thus, in both cases, shifting the real data points to the left in 
Fig.~\ref{ewt}, bottom right panel. The Si{\sc i} line is slightly 
overestimated, because it is contaminated, mainly by CO, due to the intrinsic 
velocity dispersion of the galaxies, together with the medium-resolution of 
our spectra. In conclusion, the majority of spirals are closer to giants and 
supergiants than the ellipticals, demonstrating the difference in their 
stellar populations.

Almost all spirals and ellipticals have EW(CO 2.29) $>$ 15 $\AA$ which is
 larger 
than the EW of K giants (from 10 $\AA$  to 15 $\AA$) or main sequence stars. 
In fact, the only stars with EW of CO(2.29) higher than 15 $\AA$ are M giants 
and K and M supergiants. 
The stellar population of these galaxies must 
therefore be dominated by very cool stars, probably M giants, to match the 
measured EWs. Note, however, that since M stars are much brighter in the 
$K$-band than K stars, this does not imply that M giants are the only 
significant luminosity-weighted population.

\begin{figure}
\begin{center}
\includegraphics[width=0.49\textwidth]{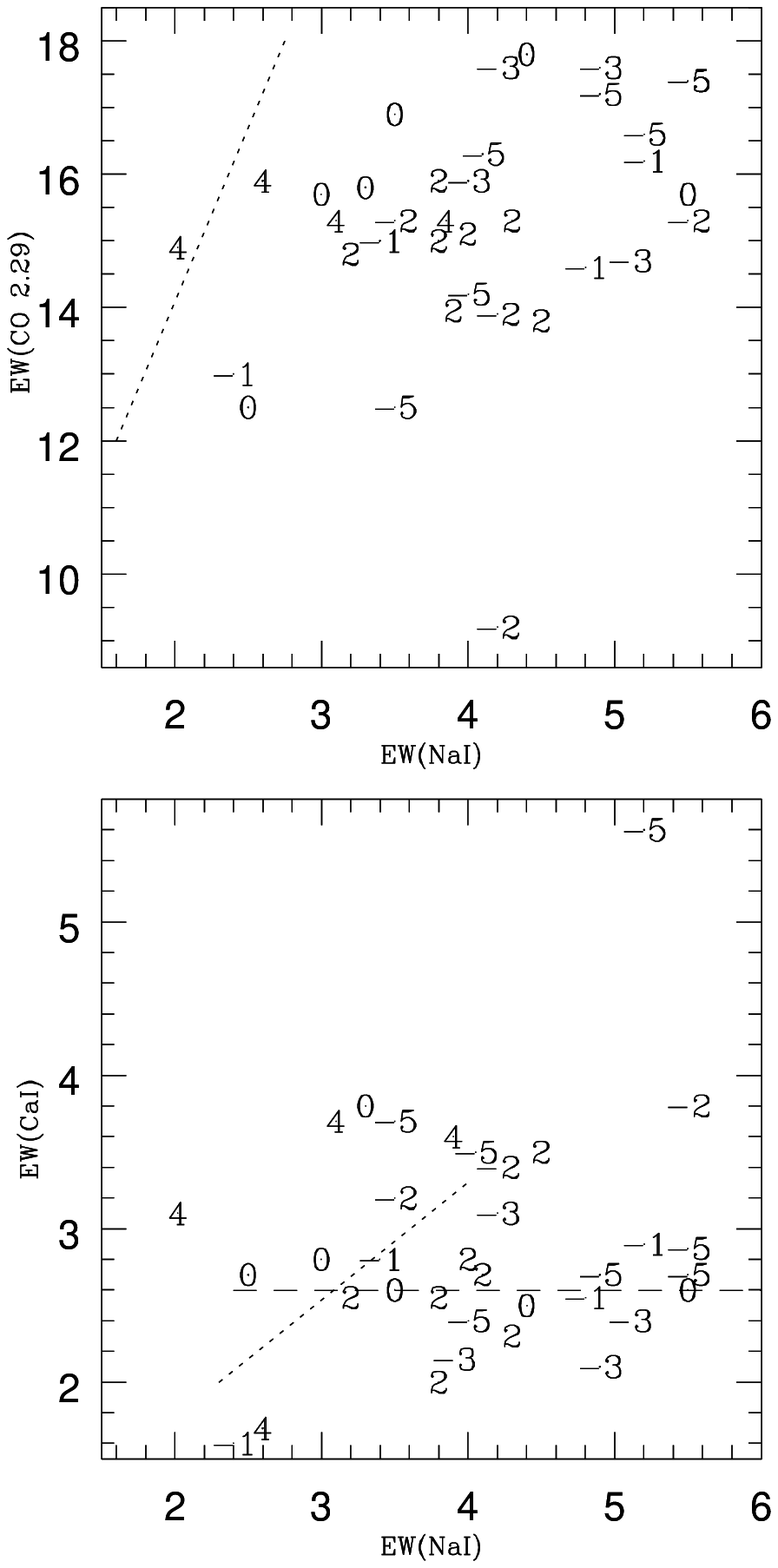}
\caption{
{\bf Top:} EW(CO 2.29) vs. EW(Na{\sc i}). 
{\bf Bottom:} EW(Ca{\sc i}) vs. EW(Na{\sc i}). 
The symbols 
are the same as in Fig.~\ref{ewt}. 
The dotted and long-dashed lines 
represent the relations of the solar metallicity cluster star fits and purely 
old population (Silva {et~al.} 2008), respectively.
\label{ewt2}}
\end{center}
\end{figure}

The EWs of CO 2.29 $\mu$m and Ca{\sc i} of the spiral and elliptical galaxies 
are plotted against EW(Na{\sc i} ) in Fig.~\ref{ewt2}. 
Majority of the spirals and ellipticals occupy a locus which is consistent 
with the relation of purely old galaxies (Silva {et~al.} 2008). 
They also have larger Na{\sc i} indices than those of solar metallicity 
cluster stars, consistent with early-type galaxies in the Fornax cluster
(Silva {et~al.} 2008).
Unlike the sample of Cesetti {et~al.} (2009), our spirals and ellipticals
do not show a significant correlation between EW(Ca{\sc i}) and EW(Na{\sc i}), 
probably due to the smaller abundance range. 
We also note that Ca and Na are 
produced by both high and low mass stars (e.g. Denisenkov \& Ivanov 1987; 
Denissenkov 2005; Langer {et~al.} 1993; Cavallo {et~al.} 1996), so that yields 
and chemical enrichment history may be relevant. 

Different star formation histories within a galaxy may set up a radial 
stellar population gradient, which can be seen as a difference in 
line strengths and EWs of absorption lines between nuclear and 
off-nuclear spectra. In non-active galaxies, EWs usually increase toward 
the bulge where they remain roughly constant. 
In the sample considered here, only a few galaxies show any indication of 
systematic radial 
variation in the strength of the lines between off-nuclear and nuclear spectra. 
While these could be due to a stellar population gradient, 
deeper and higher spatial resolution spectra are needed to confirm 
the possible gradient.


\subsection{Template spectra}

The spectra of the inactive spiral galaxies indicate that the NIR emission 
from these objects is dominated by old red stars that dominate the central 
bulges of these galaxies. By combining each individual redshift-corrected 
spectra together, we can create composite quiescent spectra for each Hubble 
type. These composite spectra can then be treated as templates representing an 
old red stellar component in the spectra of galaxies where either continuum 
emission from an AGN, hot dust, emission lines from young stars, or supernovae 
are present. These spectra therefore find applications as template spectra for 
the underlying galaxy which can be subtracted from the spectra of composite 
stellar systems (e.g. AGN host galaxies).

\begin{figure*}
\begin{center}
\includegraphics[angle=0,width=0.98\textwidth]{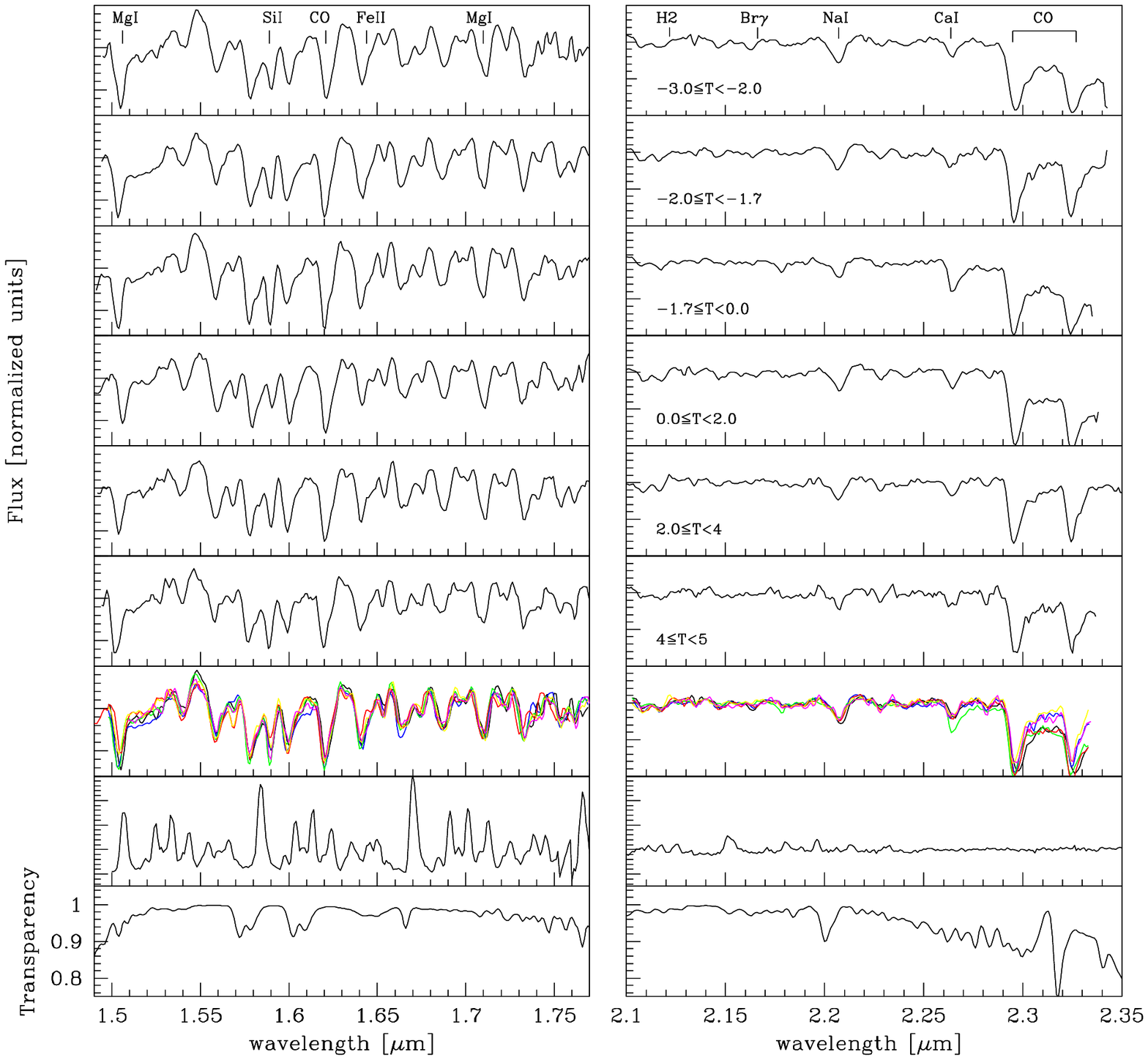}
\caption{
The $H$-band (left-hand panels) and the $K$-band (right-hand panels) 
composite nuclear spectra for the spiral galaxies, ordered by decreasing 
Hubble type from top to bottom. 
The Hubble type is listed in the $K$-band panels.
The locations of the most prominent diagnostic 
spectral absorption features together with the expected locations of main 
emission lines ([FeII], H$_2$ and Br$\gamma$) are marked. 
The panels in the third row from bottom show all the composite spectra 
grouped together, colour coded as in Fig.\ref{ewt}, to highlight their 
similarity across the Hubble type.
The two bottom panels show the sky emission and the sky transmission spectra, 
respectively.
}
\label{comb_spectra}
\end{center}
\end{figure*}

To construct the quiescent template spectra, we divided the sample into six 
groups according to their Hubble type. We first redshift-corrected all 
the spectra to the rest frame. Next, we resampled the spectra using splines, 
normalized the spectra and averaged them using the same method used to 
normalize and combine the spectra for individual galaxies. 
The resulting composite quiescent spectra, with the most prominent metal 
absorption lines and the CO band heads identified, 
are presented in Fig.~\ref{comb_spectra}, 
where we also show the sky emission and the sky transmission spectra, to give 
the reader an idea of the reliability of the template spectra.

The differences between the spectra of the galaxies of the same class are 
due both to observational uncertainties and to intrinsic differences between 
the observed galaxies. Mannucci {et~al.} (2001) and Bendo \& Joseph (2004) 
found inactive galaxies to show a high degree of uniformity, which we confirm 
with our higher spatial resolution data. 

\begin{figure}
\begin{center}
\includegraphics[angle=0,width=0.49\textwidth]{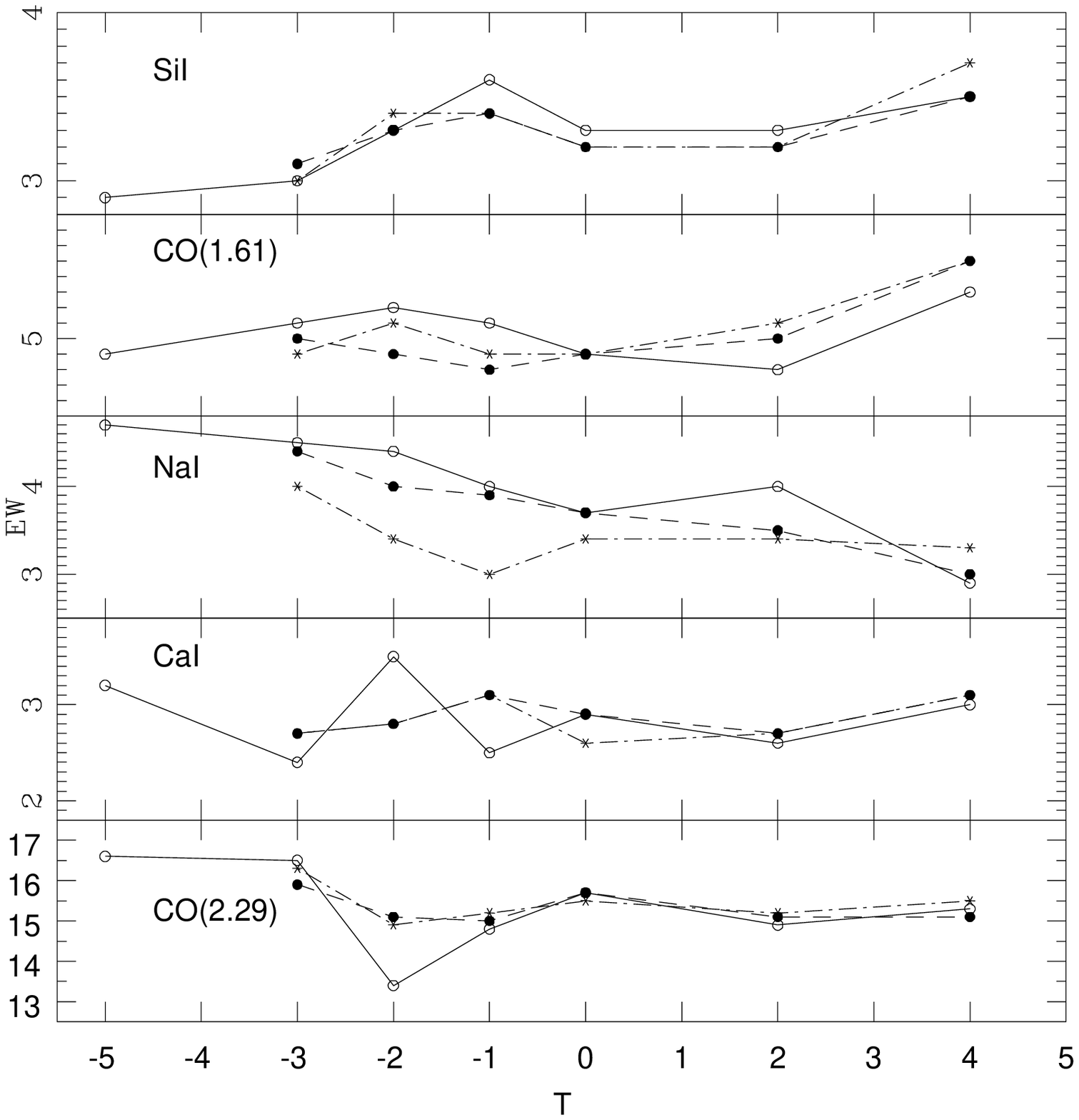}
\caption{
The average EWs of the main absorption lines as a function of the Hubble 
type T, for the nuclear spectra (solid lines), and for the off-nuclear spectra 
(off1 and off2, long-dash and dot-dash lines, respectively). 
The average errors in the average EWs are $\sim$0.5 \AA ~for the metal lines and the CO(1.62), and $\sim$1.0 \AA ~for the CO(2.29).
}
\label{ew_vs_T}
\end{center}
\end{figure}

Fig.~\ref{ew_vs_T} shows the average EWs of the main absorption lines as 
a function of the Hubble type T. We find a slight trend for the EWs of 
Si{\sc i} 1.589\,$\mu$m and CO 1.619\,$\mu$m to increase, and the EW of 
Na{\sc i} 2.207\,$\mu$m to decrease, with increasing T, but the trends are not 
statistically significant. 

\subsection{Spiral galaxies with nuclear and off-nuclear emission lines}
\label{sec:emission}

 
The nuclear NIR spectra of three galaxies (all Sb spirals) in our sample 
(NGC 2566, NGC 2935 and NGC 5612) exhibit the H$_2$ 2.122\,$\mu$m 
emission line. 
NGC 2935 is an interesting case with red nuclear colour, indicating possible 
nuclear activity. However, 
the strength of its absorption lines closely 
matches the quiescent composite spectra derived above. This implies that any 
recent star formation activity (or obscured AGN activity) lies deeply buried 
within an older population of evolved stars.

Weak emission from Br$\gamma$, the strongest hydrogen recombination line in 
the $HK$-bands, was detectable in eight sample galaxies. 
Although ionized gas may be present, the negligible recombination line flux 
in the majority of the sample implies relatively few young stars, 
and therefore relatively little recent star formation activity. 
Of the eight galaxies with detectable Br$\gamma$, five are classified in RC3 
as Sb spirals or later types (T $\geq$ 2), and none are earlier than late S0 
galaxies (T $\leq$ -2). This agrees well with earlier results presented in 
Mannucci et al. (2001) and Bendo \& Joseph (2004), and 
supports the paradigm that the relative 
strength of star formation increases along the Hubble sequence, 
probably reflecting the change along the Hubble sequence in the relative 
dominance of bulge stars to the $H$- and $K$-band emission. 


Galaxies IC 4214, NGC 2196, NGC 2775, NGC 3715, NGC 3892 and NGC 5806 exhibit 
Br$\gamma$ line but nondetectable H$_2$ line. The recombination line indicates 
that young ionizing stars are present, but the lack of shock excitation lines 
suggests that the stellar population has not evolved to a point where 
supernovae are generating shocks in the interstellar medium.  Therefore, 
using Starburst99 models (Leitherer {et~al.} 1999), 
the stellar population must be less than 3.5~Myr old.
On the other hand, NGC 5612 exhibits H$_2$ line from shock excitation 
but very weak or nonexistent Br$\gamma$ emission.  The absence of strong 
recombination lines indicates that they have at least evolved past 8~Myr.  
If the shock excitation lines are interpreted as originating from supernovae, 
then we conclude that the stellar populations in these galaxies are in 
the 8 - 36~Myr age range. 

\section{Conclusions}\label{sec:conclusions}

We have presented medium resolution NIR $HK$-band spectra of a sizable sample 
of low redshift inactive spiral galaxies, to study their stellar populations 
along the Hubble sequence, based on the diagnostic stellar absorption lines. 
Our main results are: 
(i) the EWs of the Si{\sc i}, CO(1.62) Na{\sc i}, Ca{\sc i} and CO(2.29) 
features in the spirals are most consistent with those of giant stars, 
whereas ellipticals (from literature) show a contribution from main sequence 
dwarf stars, and on average, EW(CO2.29) of spirals is somewhat greater than 
that of ellipticals. Most likely, the EWs are not significantly affected by 
non-stellar dilution.
(ii) Overall, evolved red stars completely dominate the NIR spectra, 
while contribution from hot young stars is virtually nonexistent.
(iii) We produce high spatial resolution NIR $HK$-band template spectra for 
spirals along the Hubble sequence, that will form a baseline for comparison to 
galaxies with more exotic star formation or AGN activity. 

\section*{Acknowledgments}

We thank the anonymous referee for constructive comments and suggestions. 
We acknowledge financial support from the Academy of Finland, projects 
8201017, 8107775 and 2600021611 (JK, TH) and 2600021711 (JR).
This research has made use of the NASA/IPAC Extragalactic Database (NED), 
which is operated by the Jet Propulsion Laboratory, California Institute 
of Technology, under contract with the National Aeronautics and Space 
Administration. We are grateful for the ESO staff for their support
during the observations.

\appendix

\bsp

\label{lastpage}


\begin{thebibliography}{}
\bibitem[{Ali {et~al.}(1995)Ali, Carr, DePoy, Frogel, \& Sellgren}]{ali95} Ali, B., Carr, J., DePoy, D., Frogel, J., \& Sellgren, K. 1995, AJ, 110, 2415
\bibitem[{Alonso-Herrero {et~al.} (2000)}]{alon00} Alonso-Herrero,A., Rieke,M.J., Rieke,G.H., Shields,J.C., 2000, ApJ 530, 688
\bibitem[{Bendo \& Joseph(2004)}]{bend04} Bendo, G. \& Joseph, R. 2004, AJ, 127, 3338
\bibitem[{Bernardi {et~al.} 1998}]{bern98} Bernardi, M., Renzini, A., da Costa, L. N., {et~al.} 1998, ApJ, 508, L143
\bibitem[{Bernardi {et~al.} 2002}]{bern02} Bernardi, M., Alonso, M.V., da Costa, L.N., {et~al.} 2002, AJ, 123, 2990
\bibitem[{Bettoni {et~al.}(1991)Bettoni Galletta \& Oosterloo}]{beto1991} Bettoni, D., Galletta, G., \& Oosterloo, T. 1991, MNRAS,248, 544
\bibitem[{Beuing {et~al.}(2002)Beuing {et~al.}}]{beu2002} Beuing, J., Bender, R., Mendes de Oliveira, C., Thomas, D., \& Maraston, C. 2002, A\&A, 395, 431
\bibitem[{Burston {et~al.}(2001)Burston {et~al.}}]{burs01} Burston,A.J., Ward,M.J., Davies,R.I., 2001, MNRAS 326, 403
\bibitem[{Carollo {et~al.}(1993)Carollo, Danziger \& Buson}]{car1993} Carollo, C.M., Danziger, I.J., \& Buson, L. 1993, MNRAS, 265, 553
\bibitem[{Cavallo {et~al.}(1996)Cavallo {et~al.}}]{cava96} Cavallo, R. M., Sweigart, A. V., \& Bell, R. A. 1996, ApJ, 464, L79
\bibitem[{Cesetti {et~al.}(2009)Cesetti, Ivanov, Morelli, Pizzella, Buson, Corsini, Dalla~Bonta, Stiavelli, \& Bertola}]{cese08} Cesetti, M., Ivanov, V., Morelli, L., {et~al.} 2009, A\&A, 497, 41
\bibitem[{Coziol {et~al.}(2001)Coziol {et~al.}}]{cozi01} Coziol,R., Doyon,R., Demers,S., 2001, MNRAS 325, 1081 
\bibitem[{Dalle Ore {et~al.}(1991})Dalle Ore {et~al.}]{dalle91} Dalle Ore, C., Faber, S.M., Jesus, J., Stoughton, R., \& Burstein, D. 1991, ApJ, 366, 38
\bibitem[{Dannerbauer {et~al.} (2005)}]{dann05}Dannerbauer,H., Rigopoulou,D., Lutz,D. {et~al.}, 2005, A\&A 441, 999  
\bibitem[{Davies {et~al.}(1987})Davies {et~al.}]{davies87} Davies, R.L., Burstein, D., Dressler, A., {et~al.} 1987, ApJS, 64, 581
\bibitem[{de Souza {et~al.}(1993)de Souza, Barbuy, \& Dos Anjos}]{desou1993} de Souza, R.E., Barbuy, B., \& Dos Anjos, S. 1993, AJ, 105, 1737
\bibitem[{Dressler \& Sandage}(1983)]{dres1983} Dressler, A., \& Sandage, A. 1983, ApJ, 265, 664
\bibitem[{Denisenkov \& Ivanov}(1987)]{deni87} Denisenkov, P. A., \& Ivanov, V. V. 1987, Soviet Astron. Lett., 13, 214
\bibitem[{Denissenkov} (2005)]{deni05} Denissenkov, P. A. 2005, ApJ, 622, 1058
\bibitem[{Engelbracht {et~al.} (1998)}]{eng98}Engelbracht, C.W., Rieke, M.J., Rieke, G.H., Kelly, D.M., \& Achtermann, J.M. 1998, ApJ, 505, 639
\bibitem[{Faber {et~al.}(1989)Faber {et~al.}}]{faber89} Faber, S.M., Wegner, G., Burstein, D.,{et~al.} 1989, ApJS, 69, 763
\bibitem[{Fremaux {et~al.}(2007)Fremaux, Pelat, Boisson, \& Joly}]{frem07} Fremaux, J., Pelat, D., Boisson, C., \& Joly, M. 2007, A\&A, 473, 377
\bibitem[{Frogel {et~al.}(2001)Frogel, Stephens, Ram{\'{\i}}rez, \& DePoy}]{frog01} Frogel, J.~A., Stephens, A., Ram{\'{\i}}rez, S., \& DePoy, D.~L. 2001, AJ, 122, 1896
\bibitem[{F{\"orster Schreiber}(2000)}]{fors00} F{\"orster Schreiber}, N.~M. 2000, AJ, 120, 377
\bibitem[{Goldader {et~al.}}(1997)]{gold97} Goldader,J.D., Joseph,R.D., Doyon,R., Sanders,D.B., 1997, ApJ 474, 104
\bibitem[{Golev \& Prugniel}(1998)]{gol98} Golev, V., \& Prugniel, Ph., 1998, A\&A, 132, 255
\bibitem[{Hamuy {et~al.}}(2000)]{ham00} Hamuy, M., Trager, S.C., Pinto, P.A. {et~al.} 2000, AJ, 120, 1479
\bibitem[{H\'eraudeau {et~al.}(1999}H\'eraudeau]{herau99} H\'eraudeau, Ph., Simien, F., Maubon, G., \& Prugniel, Ph. 1999, A\&AS, 136, 509
\bibitem[{Hyv\"onen {et~al.}(2009}Hyv\"onen, Kotilainen, Reunanen \& Falomo]{hyvo09} Hyv\"onen, T., Kotilainen, J., Reunanen,J., Falomo, R.. 2009, A\&A, 499, 417
\bibitem[{Idiart {et~al.}(1996})Idiart, de Freitas Pacheco \& Costa]{idia96} Idiart, T.P., de Freitas Pacheco, J.A., \& Costa, R.D.D. 1996, AJ, 112, 2541
\bibitem[{Ivanov {et~al.}(2000})Ivanov {et~al.}]{ivan00} Ivanov,V.D., Rieke, G.H., Groppi, C.E. {et~al.} 2000, ApJ, 545, 190
\bibitem[{Ivanov {et~al.}(2004})Ivanov {et~al.}]{ivan04} Ivanov,V.D., Rieke, M.J., Engelbracht, C.W. {et~al.} 2004, ApJS, 151, 387
\bibitem[{Jablonka {et~al.}(1996})Jablonka, Martin \& Arimoto]{jabl1996} Jablonka, P., Martin, P., \& Arimoto, N. 1996, AJ, 112, 1415
\bibitem[{Kleinmann \& Hall(1986)}]{klei86} Kleinmann, S. \& Hall, D. 1986, ApJS, 62, 501
\bibitem[{Lan\c{c}on {et~al.}(1999})Lan\c{c}on {et~al.}]{lanc99} Lan\c{c}on,A., Mouhcine,M., Fioc,M., Silva,D., 1999, A\&A 344, L21
\bibitem[{Langer {et~al.}(1993})Langer {et~al.}]{lark98} 	Langer, G. E., Hoffman, R., \& Sneden, C. 1993, PASP, 105, 301
\bibitem[{Larkin {et~al.}(1998})Larkin {et~al.}]{lark98} 	Larkin,J.E., Armus,L., Knop,R.A., Soifer,B.T., \& Matthews,K., 1998, ApJS 114, 59
\bibitem[{Leitherer {et~al.}(1999})Leitherer {et~al.}]{leit99} 	Leitherer,C., Schaerer,D., Goldader,J.D., et al, 1999, ApJS, 123, 3
\bibitem[{Mannucci {et~al.}(2001)Mannucci, Basile, Poggianti, Cimatti, Daddi, Pozetti, \& Vanzi}]{mann01} Mannucci, F., Basile, F., Poggianti, B., {et~al.} 2001, MNRAS, 326, 745
\bibitem[{Mannucci {et~al.}(2003)Mannucci {et~al.}}]{mann03} Mannucci, F., Maiolino, R., Cresci, G., {et~al.} 2003, A\&A, 401, 519
\bibitem[{Maraston(2005)}]{mara05} Maraston, C. 2005, MNRAS, 362, 799
\bibitem[{Mayya(1997)}]{mayy97} Mayya, Y. 1997, ApJ, 482, L149
\bibitem[{Mehlert {et~al.} 2003}]{mehl03} Mehlert, D., Thomas, D., Saglia, R. P., Bender, R., Wegner, G. 2003, A\&A, 407, 423
\bibitem[{Merrill \& Ridgway(1979)}]{merr79} Merrill, K. \& Ridgway, S. 1979, ARA\&A, 17, 9
\bibitem[{Mobasher \& James(2000)}]{moba00} Mobasher, B. \& James, P. 2000, MNRAS, 316, 507
\bibitem[{Moorwood {et~al.}(1998)Moorwood {et~al.}}]{moor98} Moorwood, A., Cuby,J.G., Lidman,C., 1998, The Messenger 91, 9
\bibitem[{Oliva {et~al.}(1995)Oliva, Origlia, Kotilainen, \& Moorwood}]{oliv95} Oliva, E., Origlia, L., Kotilainen, J., \& Moorwood, A. 1995, A\&A, 301, 55
\bibitem[{Origlia {et~al.}(1993)Origlia, Moorwood, \& Oliva}]{orig93} Origlia, L., Moorwood, A., \& Oliva, E. 1993, A\&A, 280, 536
\bibitem[{Paturel {et~al.}(2003)Paturel {et~al.}}]{pat03} Paturel, G., Petit, C., Prugniel, Ph., {et~al.} 2003, A\&A, 412, 45
\bibitem[{Ramirez {et~al.}(1997)Ramirez, DePoy, Frogel, Sellgren, \& Blum}]{rami97} Ramirez, S., DePoy, D., Frogel, J., Sellgren, K., \& Blum, R. 1997, AJ, 113, 1411
\bibitem[{Reunanen {et~al.}(2002)Reunanen {et~al.}}]{reun02} Reunanen,J., Kotilainen,J.K., Prieto,M.A., 2002, MNRAS 331, 154
\bibitem[{Reunanen {et~al.}(2003)Reunanen {et~al.}}]{reun03} Reunanen,J., Kotilainen,J.K., Prieto,M.A., 2003, MNRAS 343, 192
\bibitem[{Reunanen {et~al.}(2007)Reunanen {et~al.}}]{reun07} Reunanen, J., Tacconi-Garman, L.E. \& Ivanov, V.D. 2007, MNRAS, 382, 951
\bibitem[{Riffel {et~al.}(2008)Riffel {et~al.}}]{riffel08} Riffel,R., Pastoriza,M.G., Rodriguez-Ardila,A., Maraston,C., 2008, MNRAS 388, 803
\bibitem[{Riffel {et~al.}(2009)Riffel {et~al.}}]{riffel09} Riffel,R., Pastoriza,M.G., Rodriguez-Ardila,A., Bonatto,C. 2009, MNRAS 400, 273
\bibitem[{Schechter (1983)Schechter}]{schec1983} Schechter, P.L. 1983, ApJS, 52, 425
\bibitem[{Silva {et~al.}(2008)Silva, Kuntschner, \& Lyubenova}]{silv08} Silva, D.~R., Kuntschner, H., \& Lyubenova, M. 2008, ApJ, 674, 194
\bibitem[{Tonry (1984)Tonry}]{Tonry84} Tonry, J. 1984, unpublished
\bibitem[{Trager {et~al.}(1998)Trager {et~al.}}]{tra98} Trager, S.C., Worthey, G., Faber, S.M., Burstein, D., Gonzalez, J.J. 1998, ApJS, 116, 1
\bibitem[{Vanzi \& Rieke (1997)Vanzi \& Rieke}]{van97} Vanzi, L., \& Rieke, G.H. 1997, ApJ, 479, 694
\bibitem[{V{\'a}zquez {et~al.}(2003)V{\'a}zquez, Carigi, \& Gonz{\'a}lez}]{vazq03} V{\'a}zquez, G.~A., Carigi, L., \& Gonz{\'a}lez, J.~J. 2003, A\&A, 400, 31
\bibitem[{Vega Beltr\'an {et~al.}(2001)Vega Beltr\'an {et~al.}}]{vega2001} Vega Beltr\'an, J.C., Pizzella, A., Corsini, E.M., {et~al.} 2001, A\&A 374, 394
\bibitem[{Wallace \& Hinkle(1996)}]{wall96} Wallace, L. \& Hinkle, K. 1996, ApJS, 107, 312
\bibitem[{Wallace \& Hinkle(1997)}]{wall97} Wallace, L. \& Hinkle, K. 1997, ApJS, 111, 445
\bibitem[{Wallace {et~al.}(2000)}]{wall2000} Wallace, L., Meyer, M.R., Hinkle, K., \& Edwards, S. 2000, ApJ, 535, 325
\bibitem[{Wegner {et~al.}(2003)}]{weig02} Wegner, G., Bernardi, M., Willmer, C.N.A., {et~al.} 2003, AJ, 126, 2268
\bibitem[{Whitmore, \& Kirshner(1981)}]{whit1981} Whitmore, B.C., \& Kirshner, R.P. 1981, ApJ, 250, 43 
\bibitem[{Whitmore, \& Malumuth(1984)}]{whit1984} Whitmore, B.C., \& Malumuth, E. 1984, unpublished
\bibitem[{Worthey (1994)}]{wort94b} Worthey, G. 1994, ApJS, 95, 107
\bibitem[{Worthey {et~al.}(1992) Worthey, Faber, \& Gonzalez}]{wort92} Worthey, G., Faber, S.M., \& Gonzalez, J.J. 1992, ApJ, 398, 69 
\bibitem[{Worthey {et~al.}(1994) Worthey, Faber, Gonzalez, \& Burstein}]{wort94a} Worthey, G., Faber, S.~M., Gonzalez, J.~J., \& Burstein, D. 1994, ApJS, 94, 687
\bibitem[{Worthey \& Ottaviani(1997)}]{wort97} Worthey, G. \& Ottaviani, D. 1997, ApJS, 111, 377
\end{thebibliography}
\end{document}